**Three year outcome of the Covered Endovascular Reconstruction of the Aortic Bifurcation (CERAB) technique for aortoiliac occlusive disease**

Kim, Taeymans[1]*; Erik, Groot Jebbink[2,4]*; Suzanne, Holewijn[2]; Jasper, M. Martens[3]; Michel, Versluis[4]; Peter, C.J.M. Goverde[1]; Michel, M.P.J. Reijnen[2]

*Authors contributed equally

[1]Department of Vascular Surgery, Vascular Clinic ZNA, Antwerp, Belgium

Departments of Surgery[2] and Radiology[3], Rijnstate Hospital, Arnhem, The Netherlands

[4]MIRA Institute for Biomedical Technology and Technical Medicine, University of Twente, The Netherlands

**Corresponding author:**

Erik Groot Jebbink, Department of Surgery, Rijnstate Hospital, Wagnerlaan 55, 6815 AD Arnhem, The Netherlands. Tel: 0031 26 3786909, e-mail: erik.grootjebbink@gmail.com

ABSTRACT

Objective: The objective of the current study was to demonstrate the three year outcome of the Covered Endovascular Reconstruction of the Aortic Bifurcation (CERAB) technique for the treatment of extensive aortoiliac occlusive disease (AIOD).

Methods: Between February 2009 and July 2016, all patients treated with the CERAB technique for AIOD were identified in the local databases of two centers and analyzed. Demographics and lesion characteristics were scored. Follow-up (FU) consisted of clinical assessment, duplex ultrasound and ankle brachial indices (ABI). Patency rates and clinically driven target lesion revascularization (CD-TLR) were calculated by Kaplan-Meier analysis.

Results: 130 patients (69 male and 61 female) were treated of which 68% patients were diagnosed with intermittent claudication and 32% suffered from critical limb ischemia. The vast majority (89%) were TASC-II D lesions and the remaining were TASC-II B and C (both 5%). Median follow-up was 24 months (range 0-67 months). The technical success rate was 97% and 67% of cases were performed completely percutaneously. The ABI improved significantly from 0.65 ± 0.22 preoperatively to 0.88 ± 0.15 after the procedure. The 30-day minor and major complication rate was 33% and 7%. The median hospital stay was 2 days (range 1-76 days ). At 1 and 3-years FU 94% and 96% of the patients clinically improved at least 1 Rutherford category (2% and 0% unchanged, 4% and 4% worsened). Limb salvage rate at 1-year was 98% and 97% at three year follow-up. Primary, primary-assisted and secondary patency was 86%/91%/97% at 1-year, 84%/89%/97% at 2-year and 82%/87%/97% at 3-year FU. Freedom from CD-TLR was 87% at 1-year and 86% at both 2 and 3-year FU.

Conclusions: The CERAB technique is a safe and feasible technique for the treatment of extensive aortoiliac occlusive disease with good three year results regarding patency and clinical improvement.

INTRODUCTION

The last decades have shown a clear trend towards endovascular interventions as the first-line treatment strategy for aortoiliac occlusive disease (AIOD), also as a solution for complex lesions. Endovascular treatment of AIOD has been related to less morbidity and a shorter hospital stay, compared to open surgery [1]. In addition, open reconstruction is related to late complications, including incisional hernia formation. The kissing stent (KS) technique, using two stents abutting or 'kissing' in the central lumen of the distal aorta, is most commonly used when the aortic bifurcation is involved. A recent review showed that the primary patency of KS at 2-year follow-up is 79%, with 48% TASC C or D lesions treated [2]. The geometry of the KS configuration was previously identified as a risk factor for re-stenosis and thrombosis [3-5]. The cross configuration in KS influences the mismatch areas between the stent and vessel wall that, in turn, causes flow disturbances. This is thought to be the main cause of impaired patency, related to neointimal hyperplasia induced by low oscillating wall shear stress and stagnant blood flow [6]. To overcome these disadvantages and to achieve better long-term patency, our groups introduced a new technique in 2013 named the covered endovascular reconstruction of the aortic bifurcation (CERAB) technique [7]. We previously showed that this configuration is related to a superior flow geometry and more physiological flow patterns, in vitro [8, 9]. In 2015 the early results of the CERAB technique were published with 1-year primary and secondary patency rates of 87% and 95%, respectively. These were considered to be promising, particularly since they included all treated patients from two sites including the first-in-man [10]. In the present study we have evaluated three year outcomes of the same cohort and updated it with the more recently treated patients.

METHODS

**Patients**

All patients treated with the CERAB technique between February 2009 and July 2016 in two hospitals, the Rijnstate Hospital Arnhem (Netherlands) and the ZNA Vascular Clinic (Belgium) were identified and analyzed retrospectively. Human investigation review board approval was obtained for this study and patient informed consent was not required, related to the retrospective design. Patients treated for acute limb ischemia and/or with chimney configurations were excluded from the analysis [11]. Prior to endovascular treatment all patients were treated with anti-platelet therapy, statins and (supervised) walking exercise. Medical files were screened for demographic data, clinical status, using the Rutherford classification for chronic ischemia [12], complications and information on follow-up. Lesions were categorized according to the TASC–II criteria by assessing the Computed Tomographic Angiography (CT-A) scans [13, 14]. Procedural reports were used to extract information on the procedure and the stent types used. The pre-treatment runoff resistance score for the iliac outflow was calculated based on the runoff grading scheme as proposed by Rutherford et al [12]. A three point degree of stenosis scale was used to gauge the resistance, < 50% stenosis = 0, > 50% stenosis = 1 and occlusion = 2, based on duplex ultrasound. Weighting units are divided among the relative contribution to runoff; 2 for the external iliac artery (EIA) and 1 for the internal iliac artery (IIA). The maximal combined runoff score for left and right was 12 (left+ right = (2*2+2*1)+(2*2+2*1)). Follow-up was scheduled after 6 weeks, 6, 12, 24, 36, 48 and 60 months and consisted of clinical assessment and duplex ultrasound with ankle-brachial indices (ABI) measurements. Due to the low number of patients with a completed FU of 48 and 60 months FU up till 36 months is reported in the Results section.

**CERAB procedure**

Suitability for the technique was evaluated based on CT-A imaging. Details on the CERAB technique have been described before [7]. Briefly, two introducer sheaths are placed in the common femoral arteries. The lesion is crossed either endoluminally or subintimally, depending on the lesion characteristics (Fig 1A, verifying re-entrance). After pre-dilatation, a 9 Fr introducer sheath is inserted above the proximal margin of the aortic lesion. Thereafter, a 12-mm balloon expandable ePTFE covered stent (Atrium Advanta V12, Maquet Getinge, Hudson, NH, USA) is deployed in the distal aorta. The distal end of the stent is placed approximately 15 to 20 millimeters above the bifurcation to facilitate canalization. This stent is flared proximally with a balloon adapted to the native diameter of the distal aorta, typically with a diameter of 16 mm, to ensure full apposition to the aortic wall. This creates a funnel-shaped stent with a distal segment that is still 12 mm in diameter. Subsequently, two iliac covered stents, typically 8 mm, are positioned in the conic segment and simultaneously inflated (Fig. 1B). As treatment planning is always from healthy-to-healthy tissue, in some cases distal extensions are required (Fig. 1B). In these cases we try to preserve the internal iliac arteries, if patent, using a bare metal stent at these locations to prevent buttock claudication and erectile dysfunction. Post procedure, patients receive statin treatment and dual antiplatelet therapy for at least six months after which single antiplatelet and statin therapy is continued.

**Definitions**

Primary outcome of the study was the 3-year primary patency. Secondary outcome measures included assisted-primary patency, secondary patency, freedom from clinical driven target lesion revascularization (CD-TLR), technical success, clinical improvement, length of hospital stay, 30 day morbidity, mortality and secondary interventions. Patency was determined by means of duplex ultrasound (PSV<2.5). Primary patency was defined as patency that is obtained without the need for additional or secondary surgical or endovascular procedures [15]. Assisted-primary patency is defined as patency of the configuration achieved with the use of an additional or secondary surgical or endovascular procedure, as long as occlusion of the treated segment has not occurred [15]. Secondary patency was defined as the patency achieved by all procedures to recanalize an occluded CERAB configuration, preserving the configuration. Freedom from CD-TLR was defined as the time between the procedure and any revascularization based on restenosis and or occlusion and return or increase of symptoms. Technical success was achieved when blood flow was restored with less than 30% residual stenosis. Restenosis was defined as a PSV ratio > 2.5, measured proximal, in or distal to the graft on duplex ultrasound. Major and minor complications were registered up until 30 days after the procedure. Complications leading to transient impairment were scored as minor. Complications leading to permanent damage or death were scored as major [15]. Limb salvage rate was defined as all patients without above the ankle amputations.

**Statistics**

Data are presented as mean ± standard variation, unless stated otherwise. The Shapiro-Wilk test was used to test for normality. Statistical analysis was performed using SPSS 24 (IBM Corp. in Armonk, NY, USA). The non-parametric Chi-square test was used to test for differences in outcome with respect to previous vascular interventions, learning curve, intermittent claudication (IC) vs. critical limb ischemia (CLI) and impaired runoff. The paired student T-test was used to compare ABI pre- and postoperative. An ANOVA for repeated measures was used to test the difference in Rutherford category pre- and postoperative. Univariate analysis was used to identify any correlation between smoking, diabetes, hypertension, renal status, cardiac status and pulmonary status and patency. Kaplan-Meier survival analysis was used to calculate patency numbers. The Log-rank test was used to compare the resulting survival curves. Statistical significance was defined as P < .05.

RESULTS

During the study period a total of 130 patients (69 male and 61 female) were treated with the CERAB technique at the two sites. Patient characteristics are depicted in Table I and Figure 2. One patient was classified as Rutherford category 1, as he was preventively treated for an aortobifemoral prosthesis at risk for occlusion because of high grade stenosis at the proximal anastomosis. The majority of lesions (n=116, 89%) were classified as TASC-II D, the remaining were TASC-II B and C (both n=7, 5%). Seven patients had previously undergone a surgical reconstruction of the aortoiliac segment: five aorto-bi-iliac bypasses, one ileo-femoral and one femoro-femoral crossover bypass. Another 46 patients (35%) had previously undergone an endovascular intervention; 46% of these interventions were angioplasties of the common iliac artery (in 17% kissing balloon) and 37% were angioplasties with stenting of the common iliac artery (31% KS), Table I provides details on the remaining interventions. Prior to the intervention 35% (n=44) had no stenosis in one of the outflow vessels, 35% (n=44) had a stenosis and 30% (n=38) had an occlusion in one of the outflow vessels (either left or right). Further details per vessel segment are presented in Table I. Median follow-up was 23.5 months (range 0-67 months). Follow-up at 24 and 36 months was available in 56 and 37 cases, respectively.

**Procedural results**

The technical success rate was 97%. In 4 cases technical success was not obtained; in three patients no re-entry could be obtained and in one the lesion could not be passed. All technical failures occurred within the first 40 treated patients (case 20, 24, 37 and 39). Most procedures were performed by bilateral percutaneous access of the common femoral arteries (n=87, 67%). In 15% of cases (n=20) a surgical cut down of both femoral arteries was performed, combined with an endarterectomy of the common femoral artery in 65% of them (n=13). In 18% of the procedures (n=23) a percutaneous access technique was used on one side and surgical cut down was performed on the contralateral side. At the location of the cut down this was combined with an endarterectomy in 52% of the procedures. Brachial access was used in two procedures.

Three covered stents were used in 55% (n=67) of the cases, 4 in 19% (n=23), 5 in 22% (n=27) and more than 6 stents in 4% (n=5). The aortic stent was predominantly used in the dimensions 12x41 mm (n=77) or 12x61 mm (n=34). Limbs (left + right) were deployed with a diameter x length of 8x59 mm (n=190), 8x38 mm (n=30) and 6x59 mm (n=10). In a number of cases the limb was extended into the external iliac artery to reach healthy tissue (14% to the right and 15% to the left side). The mean procedure time was 152.7 ± 88 minutes and mean amount of contrast used was 122.5 ± 62.4 mL. Median hospital stay was 2 days (range 1 to 76 days); 54% of the patients stayed in hospital for 1-2 days; 29% of the patients stayed 3-5 days and 6% were admitted for longer than 5 days.

In 11% of the cases there were procedural complications, including unintentional dissection, arterial rupture or thrombosis (Table II) and all of them were solved during the initial procedure. In one patient an extensive bleeding in the left common iliac artery led to a resuscitation after cardiac arrest and subsequent prolonged ICU and hospital stay (76 days). The patient died, with an unknown cause of death, 3 months after discharge.

**Clinical outcome**

Postoperative complications are depicted in Table II. Minor complications occurred in 43 cases (33%), in 33 cases (77%) these were hematoma, ecchymosis or leg edema. A re-intervention was necessary in 2 cases with a minor complication, both requiring thrombin injection for a false aneurysm at the access site. In 8% of the treated patients major postoperative complications occurred: stent collapse in one of the limbs in three cases; in one case the CERAB was explanted and replaced with an aorto-bi-femoral graft and in two cases a kissing balloon technique was used to restore the flow lumen of the collapsed limb (Fig. 3). Another patient was re-operated for an occlusion of the femoral artery attributed to a misplaced closure device. In two cases an early thrombosis of the CERAB occurred, both 3 weeks postoperatively. Both were successfully treated; one by thrombectomy and the other by thrombolysis. Postoperative deterioration of chronic renal insufficiency occurred in one patient without the need for dialysis. This patient, treated for Rutherford category 5, died 4 months postoperative. The 30-day mortality rate was 0%.

Clinical improvement, expressed as an increase of at least one Rutherford category, at 6 weeks follow-up was 87%. No improvement was observed in 11% of the patients and in 3% it worsened (max. 1 category). The median Rutherford category changed from 3 (min: 1, max: 6) preoperative to a median of 0 (min: 0, max: 5) postoperative ($P<.05$). At 24 and 36 months the median Rutherford category was 0 (min: 0, max: 6) and 0 (min: 0, max: 6), both significantly different with respect to the preoperative staging, an overview is given in Figure 2. The ABI significantly improved from $0.65 \pm 0.22$ preoperatively to $0.88 \pm 0.15$ after the procedure ($P<.05$). At 24 and 36 months the ABI was $0.97 \pm 0.14$ and $0.99 \pm 0.14$.

In total 3% of patients underwent toe amputations, three within 30 days and one during the procedure (unrelated to the procedure). The preoperative Rutherford classification was 5 in two cases and 6 in two cases. After 3-year follow-up 4 (3%) major amputations (under or above the knee) were performed, at 6 months (n=1), 12 months (n=1) and 24 months (n=2) FU, respectively. The initial Rutherford classification was 4 in three cases and 5 in one case. Limb salvage rate at 1-year was 98% and 97% at three year follow-up. The overall survival at 1 and 3-year FU was 93% and 88%, respectively. In total 12 patients died during the 3-year follow-up.

**Patency**

The patency rates are depicted in Figure 4. The primary patency was 86% after 12 months, 84% after 24 months and 82% after 36 months. Univariate analyses showed no significant relation between primary patency and smoking, diabetes, hypertension, renal insufficiency or coronary disease. No differences in outcome were observed between the first 20 treated in each clinic and the patients treated thereafter (P=.54 and P=.28), and as such a clear learning curve effect could not be established.

Outcome between patients with and without a history of previous vascular interventions (either surgical or endovascular) was not significantly different (P=.26). More details are presented in Table III. There was no significant difference (P=.24) in primary patency for patients with an increased runoff score (score 1-12) prior to treatment and patients without (score 0). Furthermore, grouped initial Rutherford indication (IC vs. CLI) did not influence the primary patency results (P=.61).

**Reinterventions**

During the follow-up period CERAB-related reinterventions were performed in 18 patients (14%) (Table IV). 72% of the patients received one or two reinterventions and the majority of them were performed within the first 12 months after the initial procedure (88%). Apart from reinterventions other (endo)vascular interventions were performed in 27 patients (21%). In 67% of the cases this was either a plain balloon angioplasty (PBA) or PBA and stenting of the outflow arteries.

DISCUSSION

In the present study we have demonstrated that patency and clinical outcome of the CERAB technique for extensive AIOD is satisfying, with three-year primary, primary-assisted and secondary patency rates of 82%, 87% and 97%, respectively. The technical success rate was 100% in the last 90 procedures, and in many of them open surgery would have been the only alternative treatment modality. Technically, there are no anatomical or morphological lesion boundaries for indication of the CERAB technique. The amount of (circular) calcification is not considered to be an exclusion criterion for the technique. In cases presenting lesions just distal to the renal arteries the chimney CERAB technique of the inferior mesenteric artery was applied (not included in the present study)[11]. However, the overall health situation of the patient has to be taken into account and in relatively fit patients open surgical repair may be preferred over the use of chimney's in the visceral arteries.

Eighty-eight percent of reinterventions were performed within the first year after treatment, which is reflected in the stable patency rates afterwards. These results stand firm amidst results that are obtained with the KS technique, with 2-year primary patency of 79% (range 58-92%) and also compared to aorto-bi-femoral grafts, with a 2-year primary patency of 93% (range 87-98%) [2, 16, 17]. It should be noted, however, that more complex lesions were treated in the current study, when compared to most results reported for KS and open repair, both 50% TASC-C&D lesions, making the comparison less reliable [2, 16]. Only the cohort of Dorigo and colleagues had a comparable distribution of the TASC categories [17].

Flow and geometry seem to be important factors in view of stent patency. In previously published papers we studied the influence of the KS and CERAB technique on radial mismatch and blood flow using in vitro modeling [8, 9]. The effects of different stent configurations on flow perturbations were investigated and these studies showed that the CERAB configuration is the most unimpaired physiologic reconstruction with only a few zones of recirculation and little fluid stasis. Lowering flow disturbances and radial mismatch, i.e. mimicking a native bifurcation, could lower thrombus formation and thereby restenosis [18].

The 30-day major complication rate of 8% is low in comparison to the 30-day major complication rate of 20% reported after aorto-bi-femoral and aorto-bi-iliac bypass procedures [19] and in line with respect to those observed after KS treatment (5.8% ± 4.4%, TASC C&D in 50% of cases) [2]. Therefore, the results of this study show that the technique is indeed a valuable alternative for these treatment options.

The relation between patency and outflow has previously been shown for KS by several studies [4, 16] while it was not observed by others [20, 21]. Therefore, the impact of impaired runoff in aortoiliac stenting remains under debate. Studies that did observe the aforementioned relation pointed out that the atherosclerotic process might be more virulent in patients with extensive PAD and impaired runoff. This underlines the importance of proper risk management using statins, antiplatelet drugs and walking exercise. However, the optimal duration of double antiplatelet therapy has never been studied in this respect, and should be the topic of further research.

First signs of patency loss are usually observed on duplex ultrasound. According to the study of Chong et al. peak systolic velocity (PSV) values above average are commonly reported during duplex follow-up in the CERAB configuration, and could require a CT-A for further assessment [22]. CT-A data often reports a patent configuration, therefore Chong and colleagues proposed a tool to predict the maximal peak systolic velocity (PSV) in the CERAB. In future studies it would be interesting to incorporate such tools and to evaluate their predictive value.

Most reinterventions were performed within the first 12 months after treatment. This emphasizes that strict follow-up in the first 12 months is paramount to detect onset of re-stenosis, while follow-up afterwards could be less frequent. With this FU schedule a secondary patency rate of 97% can be achieved at three-year follow-up.

In the present study, patency was defined according to the reporting standards, as recently described by Stoner et al. [15]. In other studies, including our first paper on the outcome after CERAB, another definition was used, that includes the presence of a significant stenosis as loss of primary patency [10]. In the current cohort there were only two cases with such a stenosis left untreated, and as such the influence on the results is only minimal with a 3-year primary patency of 81% vs. 82%. Standardization of definitions and adherence to reporting standards is key and requires attention when comparing data from different studies.

In the current cohort four technical failures were reported and in all cases the lesion could not be passed from a retrograde direction. Brachial access was not attempted in these cases. Nowadays, brachial access is applied in our centers in challenging cases and is a viable option in case of a failed retrograde approach [23]. When comparing the major complication rate with the complications observed in our early series, a 4-fold increase was noted (2% vs 8%). This could be related to the fact that over time more complex lesions were considered to be suitable for the technique and also underlines frailty of this group of patients. Three of the complications were caused by a crushed iliac stent in the aortic cuff (two examples shown in Fig. 3). In one of them this was likely to be a technical error where the aortic cuff was post-dilated after positioning of the limbs, thereby crushing the contralateral limb. In the other two cases no clear reason was observed, however, could have been related to the heavily calcified lesions. Improvement of stent design and additional knowledge on stent mechanics may also contribute to the prevention of this complication. In addition, a relining stent could be used to add radial force to the CERAB configuration. However, relining the CERAB configuration as a preventive measure would significantly increase cost. Furthermore, to our knowledge no lesion morphology or geometry characteristics can predict stent crushing. When diagnosing stent collapse, PBA in combination with relining could be beneficial, since the original configuration might be weakened.

The choice of stents was based on studies that reported a superior patency of covered stents over bare metal stents in treatment of TASC C and D lesions [24-27]. In 45% of the cases we could not complete the procedure with three stents only, emphasizing the need for longer 8 mm diameter limbs (currently limited to 58 mm length). In the current cohort the CERAB was mostly constructed using Advanta Atrium V12 stents. Due to manufacturing problems since 2015 the large diameter Atrium Maquet V12 stents are no longer available. In search of alternatives the LifeStream (BARD Peripheral Vascular, Tempe, AZ, USA) and more recently the Begraft and Begraft Aortic (Bentley Innomed GmbH, Hechingen, Germany) were used for this indication. Future research should point out what the differences are between these three balloon-expandable stents with respect to placement accuracy, radial force, durability of the ePTFE layer and patency. Parallel to the introduction of the CERAB the use of the AFX unibody bifurcation endograft (Endologix Inc, Irvine, Calif, USA) was described for treatment of AIOD [28, 29]. The benefit of this stent is the fact that radial mismatch is completely prevented, but being a self-expanding stent the AFX stent has less radial force.

The present study is limited by the fact that it describes the first experience of a novel technique. With gaining experience results were likely to improve, but broadening indication for the technique during time may also have affected results. Furthermore, the retrospective analyses of our registry limited us to draw firm conclusions on causation of failure of patency. Moreover, the 3-year follow-up was not completed for the entire cohort, and as such only reflects early experience. Nevertheless, the standard error of the patency rate estimate was below 10%.

**Conclusion**

This study has shown that the CERAB technique is related to good three year results, with regard to patency and clinical outcome, in patients treated for mostly TASC-II D lesions. Long-term studies should establish its role in these patients and risk-prediction tools need to be assessed.